\RequirePackage{fix-cm}
\documentclass[smallextended]{svjour3}       % onecolumn (second format)
\smartqed  % flush right qed marks, e.g. at end of proof
\usepackage{graphicx}
\usepackage[cp1251]{inputenc}
\usepackage{amsmath}
\usepackage{amssymb}
\usepackage{cite}
\usepackage{textcomp}
\usepackage{float}
\usepackage[dvipsnames]{xcolor}
\usepackage{soul}
\usepackage{ulem} 		% for strikethough
\usepackage{upgreek}
\begin{document}

\title{Helium surface fluctuations investigated with superconducting coplanar waveguide resonator}
\titlerunning{Helium surface fluctuations investigated with superconducting CPWR} 

\author{N.R.~Beysengulov \and C.A.~Mikolas \and J.M.~Kitzman \and J.R.~Lane \and D.~Edmunds \and D.G.~Rees \and E.A.~Henriksen \and S.A.~Lyon \and J.~Pollanen}

\institute{N.R.~Beysengulov, C.A.~Mikolas, J.R.~Lane, J.M.~Kitzman, D.~Edmunds, J.~Pollanen \at Department of Physics and Astronomy, Michigan State University, East Lansing, Michigan 48824-2320, USA \\
\email{beysengu@msu.edu}  
\and D.G.~Rees, S.A.~Lyon \at EeroQ Corporation, Lansing, Michigan 48906, USA \\
\and E.A.~Henriksen \at Department of Physics, Washington University in St. Louis, St. Louis, Missouri 63130, USA
}

\date{Received: date / Accepted: date}
\maketitle

\begin{abstract}
Recent experiments on the coupling of the in-plane motional state of electrons floating on the surface of liquid helium to a microwave resonator have revealed the importance of helium surface fluctuations to the coherence of this motion. Here we investigate these surface fluctuations by studying the resonance properties of a superconducting coplanar waveguide (CPW) resonator filled with superfluid helium, where a significant fraction of the resonator's electromagnetic mode volume is coupled to the surface dynamics of the liquid. We present preliminary results on real-time CPW resonator frequency shifts driven by helium fluctuations, which are quantified via their power spectral density and compared with measurements using a commercial accelerometer. We find that a considerable contribution to the CPW resonator noise originates from the mechanical vibrations of the helium surface generated by the pulse tube (PT) cryocooler on the cryostat on which the experiments were performed. 
\end{abstract}

\section{Introduction}
The design of low temperature experiments with low mechanical vibrations/noise have recently received renewed attention. This is largely driven by the increasing popularity of cryogen-free ``dry'' systems instead of conventional ``wet'' cryostats containing a liquid helium bath volume. Much of this transition to ``dry'' cryostats has been driven by increasing operating expense produced by continuous helium consumption and interruptions in experiments for refilling the dewar with helium. In contrast, ``dry'' dilution refrigerators provide excellent continuous cooling and cost-effective performance at temperatures less than 10~mK. This is achieved by replacing the helium bath with a pulse tube (PT) cryocooler, which provides cooling down to $\sim 2-4$~K. At the same time the PT stage is a major source of a mechanical and acoustic noise~\cite{tomaru2004vibration,riabzev2009vibration}, a large portion of which is generated by the rotary valve that switches the connection of PT between high pressure helium gas output lines connected to a compressor at a frequency of $1-2$~Hz. To overcome this obstacle, there is now a large effort in the scientific community to minimize the level of vibrations in these ``dry'' cryostats. For example, scanning probe microscopy require an extremely low vibrational noise environment for precision measurements and various vibration isolation techniques at low temperatures have been demonstrated~\cite{den2014atomic,pelliccione2013design,de2019vibration}. Vibration isolation of this type has also been implemented in sub-mK systems employing nuclear demagnetization cooling systems on dry cryostats~\cite{Todoshchenko2014}. Similarly, experiments on superconducting qubit systems can also be affected by electrical noise in the coaxial cables originating from mechanical vibrations~\cite{kalra2016vibration,mykkanen2016reducing}.

This problem has also emerged in recent experiments attempting to coherently control the motion of a single electron on the surface of liquid helium, where fluctuations of the liquid surface have been identified as a major source of decoherence~\cite{koolstra2019coupling}. In these experiments an electrostatic trap determines the motional frequency of the quantized electron motion and the helium surface fluctuations can create noise in the trapping potential seen by electron. The helium surface fluctuations are naturally coupled to mechanical vibrations of the cryostat, which introduce an unwanted dephasing channel into the electron's quantized motional states. In light of this, it is important to understand these surface fluctuations and their coupling to vibrations of the cryostat. In this work we investigate the noise properties of a coplanar waveguide (CPW) resonator covered with a superfluid helium film to directly measure the fluctuations of the helium surface and quantify the contribution of the PT to these fluctuations.

\section{Experimental Results \& Discussion}
\begin{figure}
\centering
\includegraphics[width=1.0\textwidth]{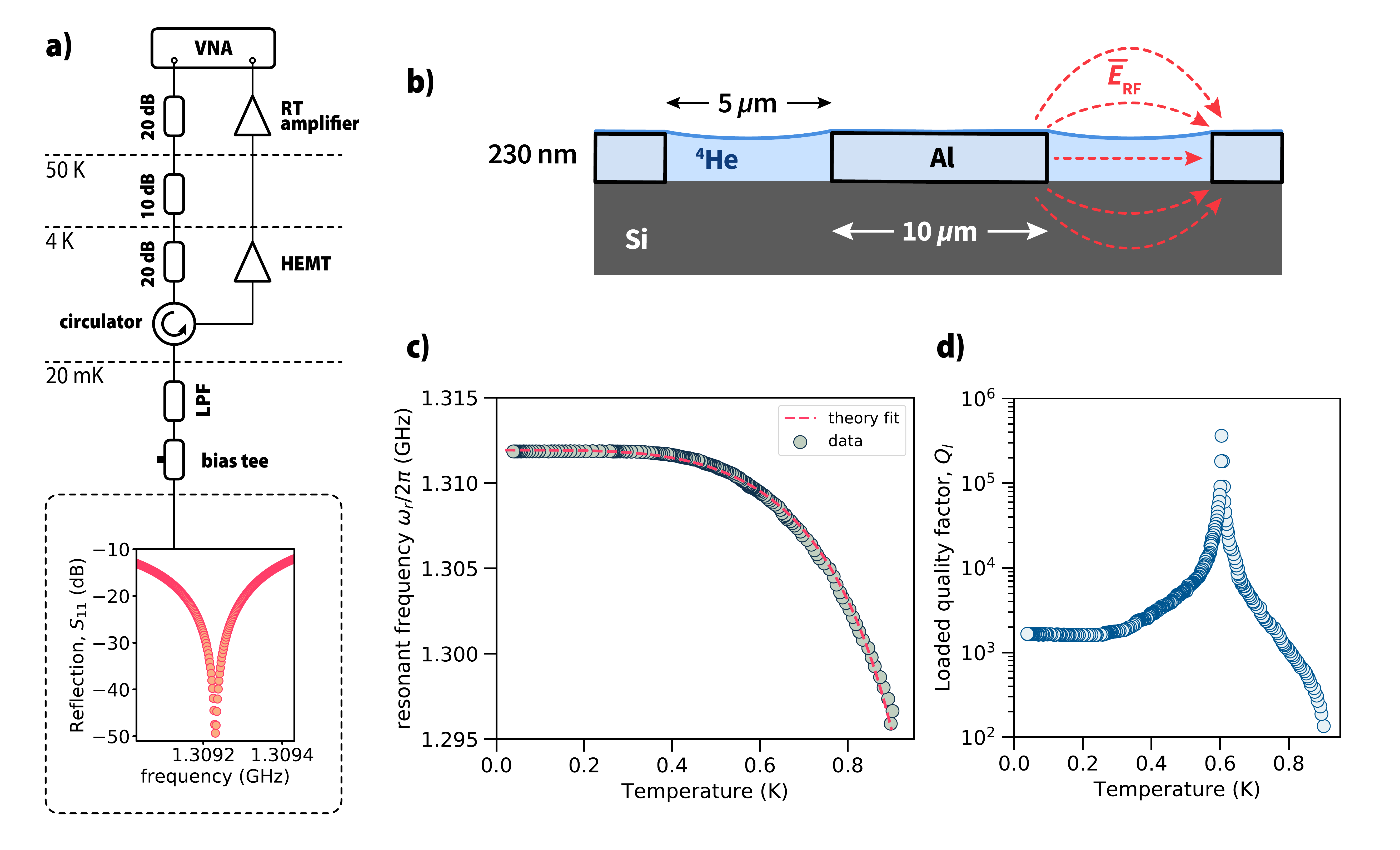}
\caption{\label{fig1} (a) The resonator measurement circuit diagram. A microwave signal generated using a vector network analyzer (VNA) is delivered into the sample cell (dashed box) through a series of cryogenic attenuators, circulator, low pass filter and bias tee (in these experiments the dc bias was not used). The reflected signal from the resonator is amplified by a cryogenic high electron mobility transistor (HEMT) amplifier and a room temperature amplifier before entering the measurement port of the VNA. The frequency dependence of $S_{11}$ acquired at $T = 0.6$~K. (b) A cross-sectional view of the resonator. Microchannels formed from the region between the resonator center strip and ground plane are filled with superfluid helium, which occupies a portion of the electromagnetic mode volume of the resonator, changing its resonance properties. (c) The temperature dependence of the measured resonance frequency (circles). The dashed line represents a fit to the theoretical model based on kinetic inductance. (d) Measured temperature dependence of the resonator loaded quality factor, which shows an anomalous increase around 0.6 K.}
\end{figure}

Figure~\ref{fig1}a and b show the schematic of our measurement setup and a cross-sectional view of the CPW resonator. A 230~nm thick aluminum (Al) $\lambda/2$-wavelength resonator was fabricated on a high-resistivity ($\rho \geq 8 $~k$\Upomega$$\cdot$cm) Si substrate. The pattern of the resonator was generated by exposing a reverse image resist AZ5214E through a mask with traditional UV lithography followed by deposition of Al via thermal evaporation. The resonator has a 10~$\mu$m wide center strip, $w = 5$~$\mu$m gaps between the center strip and the ground plane, and a length $l = 45.54$~mm corresponding to a measured resonator frequency $f_r = \omega_r/2 \pi = 1.315$~GHz. The resonator's fundamental frequency is defined by $f_r = c/2 l \sqrt{\epsilon_{\text{eff}}}$, where $c$ is the velocity of light in vacuum and the effective permittivity of the CPW line $\epsilon_{\text{eff}} = 6.25$ is close to the theoretical value $\epsilon_{\text{eff}} \approx 6$ calculated with conformal mapping techniques~\cite{goppl2008coplanar,wadell1991transmission}. The center strip of the resonator is coupled to an input transmission line via an interdigitated capacitor with six pairs of fingers of length 100~$\mu$m, width 2~$\mu$m and separation 2~$\mu$m, which gives a coupling capacitance $C_{\kappa} = 0.12$~pF based on finite element calculations.

The frequency dependence of the resonator reflection coefficient $S_{11}$ was measured with a vector network analyzer (Agilent N5230A). The input microwave signal was attenuated by 20 dB at room temperatures, followed by 10~dB and 20~dB attenuation at the 50~K and 4~K stages of the dilution refrigerator, respectively, to reduce the spectral density of thermal radiation incident on the device. This input microwave signal then propagates through a cryogenic circulator (Raditek Part no. RADC-1.35-1.45-Cryo) at the 4K stage, and a low pass filter (K\&L~4L250-4080) with a cut-off frequency of 4~GHz and a bias tee (Anritsu~K250) at the mixing chamber stage. The CPW resonator was mounted into a custom made printed-circuit board (PCB) located inside an indium o-ring sealed sample cell attached to the mixing chamber plate of the cryostat. The microwave signals enter and exit the sample cell through hermetic SMP connectors. The reflected signal from the resonator passes back through the circulator, which protects the resonator from the noise coming from the input of the cryogenic low-noise HEMT amplifier (LNF-LNC1.5\_3.5A) having a noise temperature $\simeq7$ K and a gain of 27~dB. At room temperature the reflected signal from the resonator is further amplified by a Mini-Circuits ZKL-2R5+ amplifier having a gain of 30 dB. Measurements were conducted on a BlueFors LD400 dilution refrigerator mounted on a custom tripod frame.
%, which is much smaller than the transmission line resonator capacitance $C = C_l l/2 = 3.4$~pF with $C_l = 168$~pF/m being the capacitance per unit length obtained from FEM. 

\begin{figure}
\centering
\includegraphics[width=0.75\textwidth]{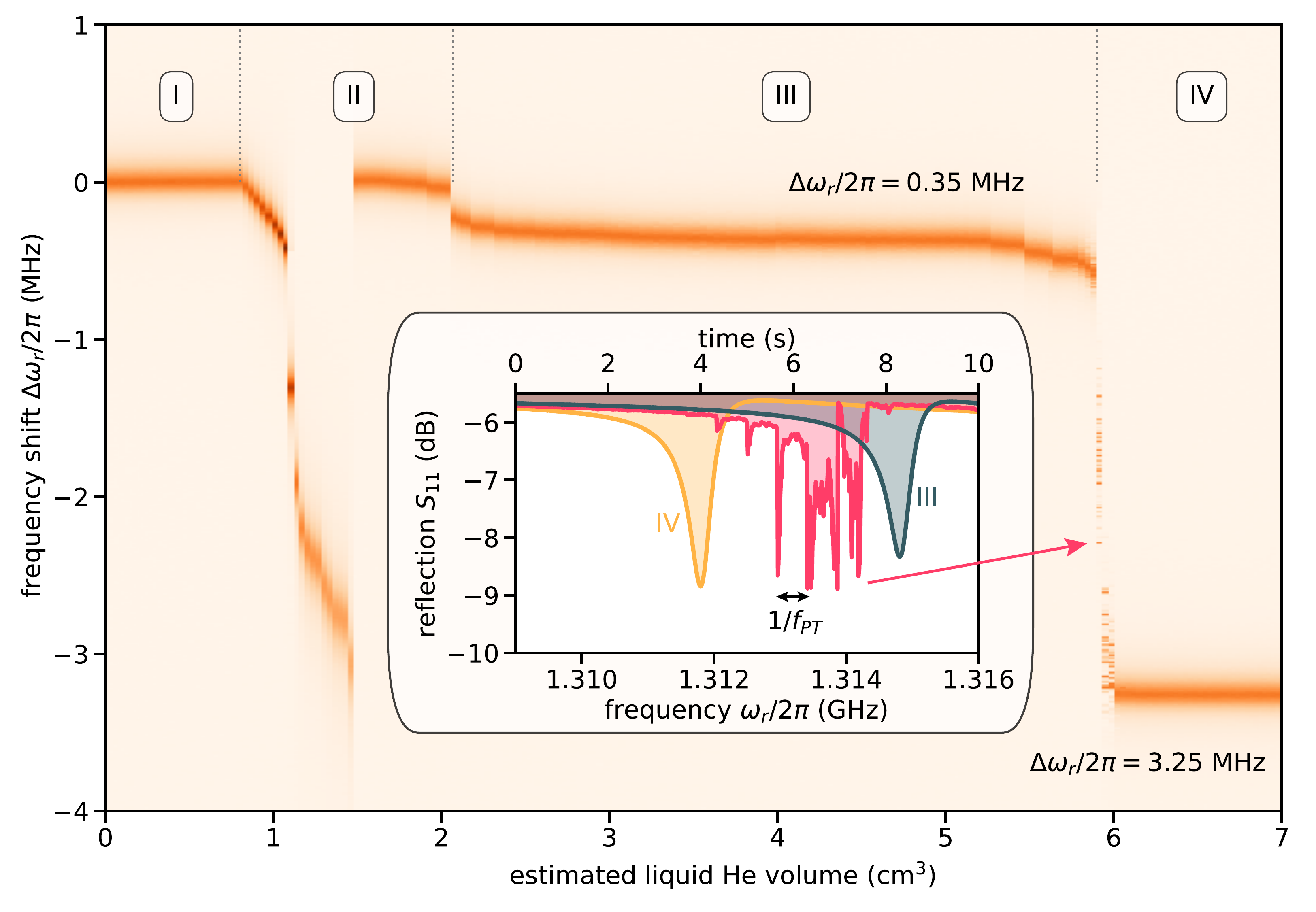}
\caption{\label{fig2} Resonator frequency shift during liquid helium condensation. Regions I, II, III and IV indicate different filling regimes as described in the main text. The inset shows vertical line cuts in regions III and IV demonstrating distinct shifts of the resonance frequency. The line cut at the boundary between regions III and IV shows the periodically distorted resonance conditions at the pulse tube frequency, which are observed during a sufficiently slow frequency sweep of the resonator.}
\end{figure}

A typical measurement of the reflection from the resonator as a function of frequency is shown in Figure~\ref{fig1}a (completely filled with helium at $T = 0.6$~K) and in the inset of Figure~\ref{fig2} (measured at $T \sim 0.1$~K with various levels of helium filling). The data were acquired with an input power of $-50$~dBm, low enough to avoid nonlinearities produced by high power excitation of the resonator. We note that the resonance curves in the inset of Figure~\ref{fig2} show small asymmetry, likely arising from impedance mismatch between the resonator and the transmission lines, which are connecting with bonding wires~\cite{megrant2012planar}. Figure~\ref{fig1}c shows the temperature dependence of the resonator's resonant frequency. In order to describe this dependence we use a lumped element approximation to model the resonator. In this model the resonance frequency scales with inductance as $\omega_r \propto 1/\sqrt{L}$. The inductance here is a sum of a temperature independent geometric (magnetic) inductance $L_m$ and the temperature dependent kinetic inductance of a superconductor $L_k$. The temperature dependence of the kinetic inductance is governed by the London penetration depth $\lambda$ and scales as $L_k \propto \lambda^2(T) \propto (1-(T/T_c)^4)^{-1}$~\cite{frunzio2005fabrication,schmidt1997physics} where $T_c = 1.2$~K  is the critical temperature for Al. We find that the best fit of this model to the experimental data is achieved with the ratio $L_k/L_m \approx 0.06$ (see Fig.~\ref{fig1}c). 

We also extract the loaded quality factor of the resonator, which is given by $Q_l = f_r/2\delta f_r$, where $\delta f_r$ is the full width at half-maximum of the resonator power spectrum. Figure~\ref{fig1}d shows the measured temperature dependence of $Q_l$. At low temperatures we find that $Q_l$ saturates at values $1.7 \times 10^3$. Interestingly, we observe an anomalous increase of the quality factor by 2 orders of magnitude near the temperature $T = 0.6$~K, which does not depend on whether the resonator covered with liquid helium or not. We believe that this behavior originates from the internal properties of the resonator, however the origin of the physics remains unknown. We note that this unexpected behavior is not critical for our discussion of helium fluctuations discussed below. However, future experiments with different geometries would be necessary to investigate the origin of this unusual temperature dependence.

We now turn our attention to the use of the resonator to investigate the surface fluctuations of the helium filling the resonator. When the vacuum above the resonator is replaced with a low-loss dielectric medium - in this case liquid helium - capacitive changes in the electrodynamic environment result in a shift of the resonator resonance frequency. The response of a CPW resonator to the introduction of liquid helium has already been investigated in Ref.~\cite{yang2016coupling,koolstra2019coupling} and \cite{koolstra2019trapping}, using sample geometries similar to ours. The authors found clear frequency shifts related to the coverage of the sample with a thin van der Waals film, followed by larger shifts due to the filling of the channels surrounding the resonator center pin. Here we observe qualitatively similar behaviour as shown in Figure~\ref{fig2}, where we monitor the frequency shift by measuring the reflection during helium condensation into the cell. The liquid helium is introduced into the sample cell through a stainless steel capillary line, which is thermally anchored using a sintered copper heat exchanger at the 1K stage of the cryostat. During the condensation process the temperature of the mixing chamber increased from 100 mK to 200 mK. We estimate the approximate amount of condensed liquid helium from the pressure change in a helium tank with known volume at room temperatures. During slow condensation we observe four regions in the response of the resonator with distinct behaviour, which are shown in Figure~\ref{fig2}. In region I the resonance frequency does not change, which we associate with the filling of the dead volume in the copper sinter by helium. Region II indicates the introduction of helium into the cell. The initial large shift of the $\omega_r/2\pi$ observed in this region is associated with an increase of the resonator's temperature due to the introduction of relatively hot helium initially thermalized at the 1K stage. After approximately 3-5 minutes the resonance frequency abruptly shifts back to nearly its original value, which we attribute to the thermalization of the liquid helium to the mixing chamber plate and the formation of an unsaturated thin van der Waals film that covers all the internal surfaces in the sample cell, including the surface of the resonator. We note that the behaviour in region II is reproducible when the cell is warmed to room temperature, fully evacuated, and subsequently cooled down and filled again. %However, if the cell is warmed up only to 20 K (which results in residual $^4$He gas in the cell) and cooled back down, this feature is not observed. 
We also note that slight changes in the helium condensation rate, which are controlled by a needle valve at room temperature, slightly change the fine scale details of the observed resonator response in region II but not the coarse scale qualitative behavior. These slight differences likely indicate nonequilibrium processes during the initial thermalization of the liquid helium and the sample substrate. 

With continued condensation a bulk helium volume is formed at the bottom of the cell, beneath the resonator, and a helium film begins to grow on the resonator. The thickness of this helium film is given by $d = (\gamma / \rho g H)^{1/4}$~\cite{pobell2007matter}, where $\gamma$ is the van der Waals constant, $\rho = 146$~kg/m$^3$ is the mass density of liquid helium, $g$ is the acceleration due to gravity, and $H$ is the distance from the resonator surface down to the bulk liquid helium level in the cell. At the same time the microchannel regions between the central pin and ground plane starts to fill with superfluid due to capillary action (region III in Fig.~\ref{fig2}). The thickness of the liquid helium in the microchannels $h$ is determined by a balance of the hydrostatic pressure caused by gravity and the surface tension of the liquid, and is given by $h \simeq d_r - \rho g H w^2/16 \sigma_t $~\cite{marty1986stability}, where $d_r = 230$~nm is the depth of the microchannel, and $\sigma_t = 3.58 \times 10^{-4}$~N/m is the surface tension of liquid helium. Once the helium fills the microchannels completely we measure the frequency shift of the resonator $\Delta \omega_r/2\pi = -0.35$~MHz, which is in good agreement with Finite Element Modeling (FEM) calculations $\Delta \omega^{\text{FEM}}_r/2\pi = -0.31$~MHz. The onset of region IV indicates that the helium level $H \rightarrow 0$, corresponding to the formation of bulk liquid above the resonator chip. Interestingly at the boundary between regions III and IV we do not observe a single resonance peak in the power spectrum, but rather a series of sharp periodic changes in the reflection coefficient appear during the frequency sweep over a 10~s time span (see inset Fig.~\ref{fig2}). The frequency of these periodic changes in the spectrum are equal to the fundamental mode of the pulse tube at 1.4~Hz. This indicates the generation of surface excitations in the bulk helium by mechanical vibrations in the cryostat driven by the pulse tube. In the last region IV we measure the frequency shift $\Delta \omega_r/2\pi = -3.25$~MHz, which is also in a good agreement with our numerical calculations $\Delta \omega^{\text{FEM}}_r/2\pi = -3.33$~MHz when the resonator is completely submerged by a thick layer of helium.
\begin{figure}
\centering
\includegraphics[width=1.0\textwidth]{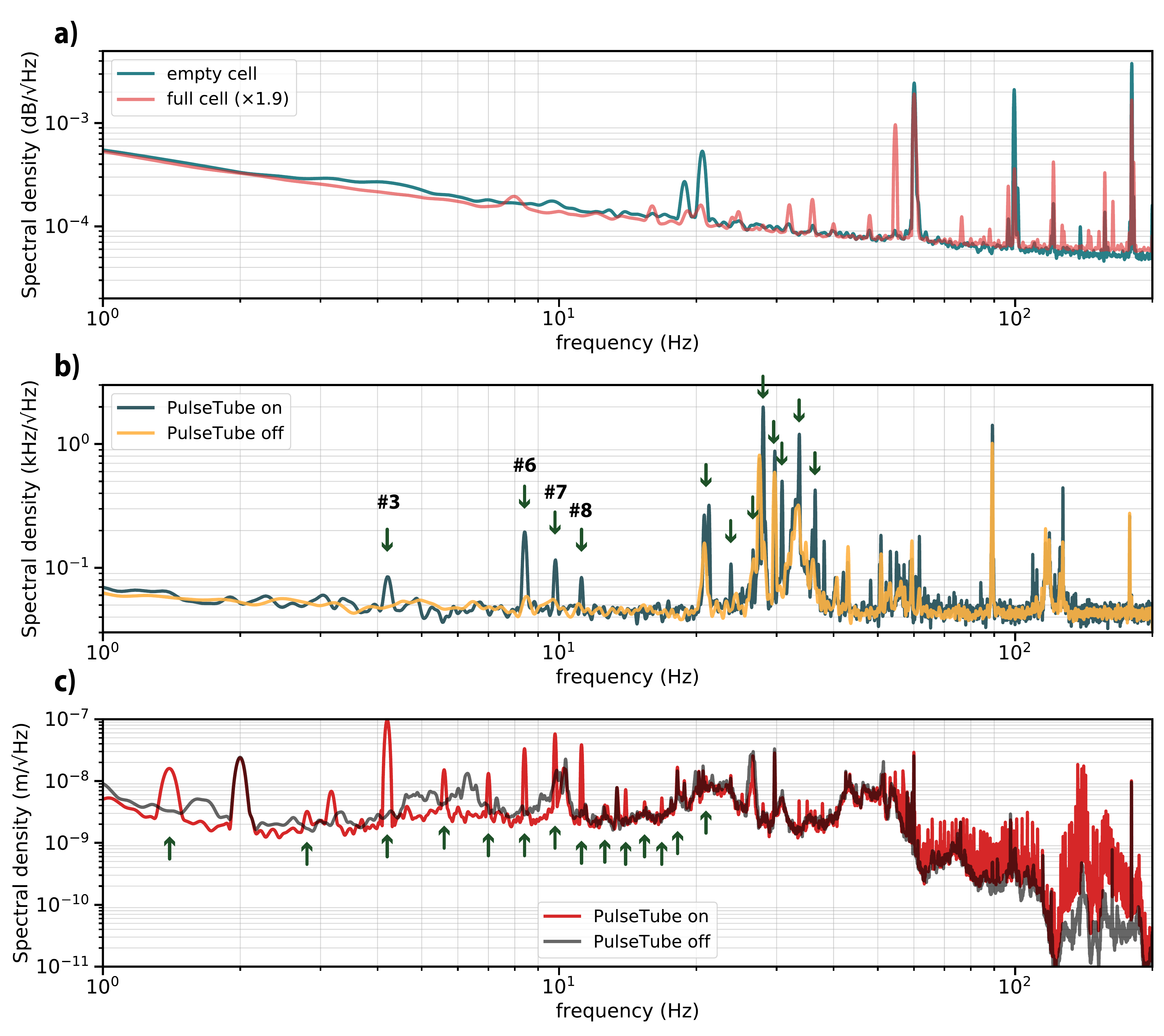}
\caption{\label{fig3} (a) Noise spectral density $S_{\text{dB}}$ of the resonator in the absence of liquid helium and when the resonator completely submerged into liquid. Measurements are done at $T = 20$~mK. The noise spectrum corresponding to the filled cell has been multiplied by a factor of 1.9 to facilitate comparison with the data taken with the cell empty (b) Spectral density $S_f$ of the frequency fluctuations with superfluid partially filling the microchannels. PT harmonics are indicated by arrows. (c) Noise spectral density of vertical vibrations of the mixing chamber plate measured via geophone. Measurements are done at room temperatures. Arrows show PT harmonics.}
\end{figure}

In order to measure helium surface fluctuations we partially fill microchannels with liquid helium and operate the device in region III shown in Figure~\ref{fig2}. The fluctuations of the helium surface change the participation volume of the dielectric medium in the electromagnetic mode volume, thus changing resonator's frequency. Therefore the information about the helium fluctuations is embedded in the resonator's noise properties. To measure this noise, we record the reflection coefficient as a function of time at fixed frequency $f_0$ corresponding to the highest slope on $S_{11}(f)$. Time traces of the reflected magnitude $S_{11}(t)$ having a duration of 40~s were recorded at a sampling rate 400 Sa/s limited by the bandwidth of the VNA. The spectral density of the measured noise $S_{\text{dB}}$ is computed using Welch's average periodogram method~\cite{welch1967use}. For reference, we show the measured noise spectral densities in the absence of liquid helium (region I) and when the resonator is completely submerged in liquid helium (region IV) in Fig.~\ref{fig3}a. Here we observe 60~Hz electric noise and several distinct resonances, potentially originating from mechanical fluctuations of the bonding wires connected to the resonator. The spectral density changes qualitatively when liquid helium fills microchannels, i.e. in region II. Figure~\ref{fig3}b shows the spectral density of the frequency fluctuations in this region calculated as $S_f = S_{\text{dB}} (|\partial S_{11}/\partial f|_{f=f_0})^{-1}$ measured at 10~mK. We observe several resonances corresponding to harmonics of the pulse tube at frequencies $n \cdot f_{\text{PT}}$, which were absent in Fig.~\ref{fig3}a. These resonances can be modestly enhanced by intentionally inducing additional mechanical vibrations in the cryostat frame. The majority of PT harmonics disappear or their amplitude reduces when the PT is turned off. A noticeable feature in the frequency range 30-60~Hz likely originates from building vibrations, which excite cryostat vibrational modes. This was confirmed by producing a frequency shift of this feature by applying additional clamping to the structure to the cryostat's outer vacuum can to modify the mechanical boundary conditions of the cryostat. These results all indicate that the majority of helium fluctuations originate from mechanical vibrations of the cryostat body and frame. The integrated RMS vibrations are calculated as:
\begin{equation}
	\Delta h_{\text{RMS}} = \Big ( \frac{\partial f_r}{\partial h} \Big )^{-1} \sqrt{\int_{f_1}^{f2} S_f^2 df},
	\label{eq:1}
\end{equation}
where $\partial f_r/ \partial h = 1.4$~kHz/nm is obtained from FEM calculations. We estimate $\Delta h_{\text{RMS}} = 0.9$~nm by integration of the data in Fig.~\ref{fig3}b between 1~Hz and 200~Hz in the case where the PT is on. This value reduces to 0.77~nm (14\% reduction) when the PT is turned off for several minutes.

To gain additional information about the cryostat vibrations we have performed measurements using a commercial geophone attached to the mixing chamber plate of the cryostat. The geophone consists of a spring loaded mass attached to a magnet, which moves relative to a solenoid. Small vertical motion of the magnet produces a voltage on the solenoid, which can be converted to a displacement. We use Geophone Sensor RTC-4.5Hz-375, which has a natural frequency of 4.5~Hz, below which the sensitivity of the sensor is small. The signal from the geophone is amplified using an NF~LI-75A low noise current preamplifier with gain 100 and recorded with a MC~USB-1602HS digitizer at a rate of 2~kSa/s. We performed a geophone calibration procedure to relate measured voltages to the magnet displacement as described in Ref.~\cite{van2005simple}. Figure~\ref{fig3}c shows the noise spectrum of the vibrations measured on the geophone attached to the mixing chamber plate at room temperature with PT on and off. The PT harmonics are clearly visible at low frequencies and at frequencies $> 60$~Hz, which disappear once the PT is off. From this measurement we estimate the total root mean square displacement noise with the PT on and off is 58~nm and 47~nm, respectively. The 19\% reduction in RMS displacement measured with geophone is in reasonable agreement with our helium filled resonator measurements. 

All the preliminary experiments, presented here, consistently indicate mechanical vibrations originating from the PT cryocooler, which then couple to helium fluctuations in the microchannels. There are several possible solutions to reduce these fluctuations of the helium surface, for example by mechanically disconnecting the cold head from the cryostat frame, or by using a spring suspended stage~\cite{de2017ultra,pelliccione2013design} and/or adding eddy current dampers~\cite{den2014atomic}. These various vibration isolation techniques should be applied with careful consideration of thermal anchoring and with feedback from measurements, such as those described here, to characterize improvements. As an alternative solution, the use of dry cryostats with ``helium batteries'' located at the 4K stage offer a compromise solution to the vibrations created by the PT. In these systems the PT can be turned off for several hours, while evaporating liquid helium in the ``helium battery'' volume provides the necessary cooling power to the 4K stage. Finally we note that the residual $\sim 80$~\% of vibrational noise potentially originates from the building and cryostat body and frame vibrations, or is the result of residual fluctuations excited by the pulse tube which ring down over longer time scales. Future experiments are needed to identify these contributions and methods to suppress them.

\section{Conclusion}
In conclusion, we have studied the properties of a CPW resonator in the presence of superfluid helium filled microchannel structures. We have identified different regimes of the resonator's response depending on the amount of liquid helium in the sample cell. Our results on the frequency shift of the resonator are consistent with FEM calculations and previous measurements. We have identified the contribution to vibrations induced by the PT, which need to be eliminated in order to achieve strong coupling between a microwave resonator and the in-plane motion of a trapped single electron.%  There remain open questions, e.g. the anomalous temperature dependence of the quality factor, which needs a further investigation.

\begin{acknowledgements}
We thank K. Nasyedkin, G. Koolstra, B. Dizdar, D.I. Schuster for useful discussions and B. Bi for use of the W.M. Keck Microfabrication Facility at MSU. This work was supported by a sponsored research grant from EeroQ Corp. Additionally, J. Pollanen, J.R. Lane and J.M. Kitzman acknowledge support from the National Science Foundation via grant number DMR-2003815 as well as the valuable support of the Cowen Family Endowment at MSU.

The data that support the findings of this study are available from the corresponding author upon reasonable request.
\end{acknowledgements}

\bibliographystyle{spphys}       % APS-like style for physics

\begin{thebibliography}{10}
\providecommand{\url}[1]{{#1}}
\providecommand{\urlprefix}{URL }
\expandafter\ifx\csname urlstyle\endcsname\relax
  \providecommand{\doi}[1]{DOI \discretionary{}{}{}#1}\else
  \providecommand{\doi}{DOI \discretionary{}{}{}\begingroup
  \urlstyle{rm}\Url}\fi

\bibitem{tomaru2004vibration}
T.~Tomaru, T.~Suzuki, T.~Haruyama, T.~Shintomi, A.~Yamamoto, T.~Koyama, R.~Li,
  Cryogenics \textbf{44}(5), 309 (2004)

\bibitem{riabzev2009vibration}
S.~Riabzev, A.~Veprik, H.~Vilenchik, N.~Pundak, Cryogenics \textbf{49}(1), 1
  (2009)

\bibitem{den2014atomic}
A.~Den~Haan, G.~Wijts, F.~Galli, O.~Usenko, G.~Van~Baarle, D.~Van Der~Zalm,
  T.~Oosterkamp, Review of Scientific Instruments \textbf{85}(3), 035112 (2014)

\bibitem{pelliccione2013design}
M.~Pelliccione, A.~Sciambi, J.~Bartel, A.~Keller, D.~Goldhaber-Gordon, Review
  of Scientific Instruments \textbf{84}(3), 033703 (2013)

\bibitem{de2019vibration}
M.~de~Wit, G.~Welker, K.~Heeck, F.M. Buters, H.J. Eerkens, G.~Koning,
  H.~van~der Meer, D.~Bouwmeester, T.H. Oosterkamp, Review of Scientific
  Instruments \textbf{90}(1), 015112 (2019)

\bibitem{Todoshchenko2014}
I.~Todoshchenko, J.P. Kaikkonen, R.~Blaauwgeers, P.J. Hakonen, A.~Savin, Review
  of Scientific Instruments \textbf{85}(8), 085106 (2014).
\newblock \doi{10.1063/1.4891619}.
\newblock \urlprefix\url{https://doi.org/10.1063/1.4891619}

\bibitem{kalra2016vibration}
R.~Kalra, A.~Laucht, J.P. Dehollain, D.~Bar, S.~Freer, S.~Simmons, J.T.
  Muhonen, A.~Morello, Review of Scientific Instruments \textbf{87}(7), 073905
  (2016)

\bibitem{mykkanen2016reducing}
E.~Mykk{\"a}nen, J.~Lehtinen, A.~Kemppinen, C.~Krause, D.~Drung,
  J.~Nissil{\"a}, A.~Manninen, Review of Scientific Instruments
  \textbf{87}(10), 105111 (2016)

\bibitem{koolstra2019coupling}
G.~Koolstra, G.~Yang, D.I. Schuster, Nature communications \textbf{10}(1), 1
  (2019)

\bibitem{goppl2008coplanar}
M.~G{\"o}ppl, A.~Fragner, M.~Baur, R.~Bianchetti, S.~Filipp, J.M. Fink, P.J.
  Leek, G.~Puebla, L.~Steffen, A.~Wallraff, Journal of Applied Physics
  \textbf{104}(11), 113904 (2008)

\bibitem{wadell1991transmission}
B.C. Wadell, \emph{Transmission line design handbook} (Artech House Microwave
  Library, 1991)

\bibitem{megrant2012planar}
A.~Megrant, C.~Neill, R.~Barends, B.~Chiaro, Y.~Chen, L.~Feigl, J.~Kelly,
  E.~Lucero, M.~Mariantoni, P.J. O'Malley, et~al., Applied Physics Letters
  \textbf{100}(11), 113510 (2012)

\bibitem{frunzio2005fabrication}
L.~Frunzio, A.~Wallraff, D.~Schuster, J.~Majer, R.~Schoelkopf, IEEE
  transactions on applied superconductivity \textbf{15}(2), 860 (2005)

\bibitem{schmidt1997physics}
V.V. Schmidt, V.~Schmidt, P.~M{\"u}ller, A.V. Ustinov, \emph{The physics of
  superconductors: Introduction to fundamentals and applications} (Springer
  Science \& Business Media, 1997)

\bibitem{yang2016coupling}
G.~Yang, A.~Fragner, G.~Koolstra, L.~Ocola, D.~Czaplewski, R.~Schoelkopf,
  D.~Schuster, Physical Review X \textbf{6}(1), 011031 (2016)

\bibitem{koolstra2019trapping}
G.~Koolstra, Trapping a single electron on superfluid helium using a
  superconducting resonator.
\newblock Ph.D. thesis, The University of Chicago (2019)

\bibitem{pobell2007matter}
F.~Pobell, \emph{Matter and methods at low temperatures}, vol.~2 (Springer,
  2007)

\bibitem{marty1986stability}
D.~Marty, Journal of Physics C: Solid State Physics \textbf{19}(30), 6097
  (1986)

\bibitem{welch1967use}
P.~Welch, IEEE Transactions on audio and electroacoustics \textbf{15}(2), 70
  (1967)

\bibitem{van2005simple}
F.~Van~Kann, J.~Winterflood, Review of scientific instruments \textbf{76}(3),
  034501 (2005)

\bibitem{de2017ultra}
L.~De~Lorenzo, K.~Schwab, Journal of Low Temperature Physics \textbf{186}(3-4),
  233 (2017)

\end{thebibliography}

%\begin{thebibliography}{}
% etc
%\end{thebibliography}

\end{document}